\documentclass[final,1p,times]{elsarticle}
\usepackage{amssymb}
\usepackage{amsmath}
\usepackage{times}
\usepackage{ascmac}
\usepackage[colorlinks=true,citecolor=blue,linkcolor=blue,urlcolor=blue]{hyperref}
\usepackage[T1]{fontenc}
\usepackage{textcomp}
\usepackage{color}
\usepackage{framed}

\definecolor{shadecolor}{gray}{0.80}

\newenvironment{myleftbar}{
  
  \MakeFramed {\advance\hsize-\width \FrameRestore}}
{\endMakeFramed}

\newcommand{\cov}{\text{cov}}
\newcommand{\cor}{\text{cor}}
\newcommand{\var}{\text{var}}

\journal{Applied Radiation and Isotopes}

\begin{document}

\begin{frontmatter}

\title{EXFOR utility codes (ForEXy) and their application to neutron fission cross section evaluation}

\author[label1]{Naohiko Otuka\corref{cor1}}
\ead{n.otsuka@iaea.org}
\affiliation[label1]{organization={Nuclear Data Section, Division of Physical and Chemical Sciences, Department of Nuclear Sciences and Applications, International Atomic Energy Agency},
%           addressline={},
            postcode={A-1400},
            city={Wien},
%           state={},
            country={Austria}}

\author[label2]{Vidya Devi}
\affiliation[label2]{organization={Department of Physics, Panjab University},
%           addressline={},
            city={Chandigarh},
            postcode={160014},
%           state={},
            country={India}}

\author[label3]{Osamu Iwamoto}
\affiliation[label3]{organization={Nuclear Data Center, Japan Atomic Energy Agency},
%           addressline={},
            city={Tokai-mura},
            postcode={319-1195},
            state={Ibaraki},
            country={Japan}}

\begin{abstract}
We developed a set of EXFOR utility codes (ForEXy) to process the information of the experimental nuclear reaction data stored in the EXFOR library.
We designed a new JSON format (J4) for the EXFOR library,
and developed a code converting the information in an EXFOR file to a J4 file (X4TOJ4) and another code converting it to an EXFOR file (J4TOX4) as a core part of this new code package.
We also developed some other codes for managements of the EXFOR storage, bibliography and dictionary.
As an application of the new code package,
we constructed covariance matrices for the fast neutron induced fission cross sections of neptunium-237 in the EXFOR library by using the new codes, and applied them to evaluation of the cross section between 100~keV and 200~MeV.
\end{abstract}


\end{frontmatter}

\section{Introduction}
\label{sec:intro}
Simulation of nuclear process is routinely done in the nuclear application field to satisfy economical and safety requirements in development and use of nuclear instruments and products.
The outputs of the simulations may strongly depend on the nuclear data adopted in the simulations.
Significant efforts have been made to improve the quality of the nuclear data, and the outcomes have been compiled in evaluated nuclear data libraries, which are further processed to meet the input specifications of simulation codes.
Nuclear reaction data evaluations are typically performed with reaction model codes.
They include many phenomenological parameters such as optical potential parameters and level densities,
and they are adjusted to describe the experimental reaction data.
The evaluated data must be therefore compared with the existing experimental reaction data to see the performance of the parameters and models.
For this purpose,
the experimental data obtained at laboratories should be always compiled with the descriptions of the experiments in a consistent manner,
and be ready for easy retrieval and extraction by data evaluators.

The EXFOR library~\cite{Otuka2014Towards} has served as the unique storage of the experimental nuclear reaction data in the world for more than 60 years.
It was initially for exchange of the neutron induced reaction data collected by the Four Neutron Data Centres,
and they developed the EXFOR format (\underline{Ex}change \underline{For}mat)~\footnote{
This abbreviation was approved in the NRDC 1970 meeting with the following statement: 
``It is noteworthy to point out that this system,
  which makes efficient use of the existing computer facilities in the four centres,
  is unique and probably the first of a kind in the field of international cooperation in the nuclear energy field~\cite{Lorenz1969Report}.
} to exchange the data accumulated in their local storage systems and formats.
Some years later,
the scope of the EXFOR library was extended to the charged-particle- and photon-induced reaction data,
and the Four Neutron Centres were reorganised to the International Network of \underline{N}uclear \underline{R}eaction \underline{D}ata \underline{C}entres (NRDC)~\cite{Otuka2014International}.

The EXFOR format was designed considering punch cards as the recording medium.
Its 80-column structure, also seen in other nuclear data formats such as the ENDF-6 format~\cite{Brown2023ENDF} and ENSDF format~\cite{Tuli2001Evaluated}, seems old-fashioned.
However, the EXFOR format is logical and human readable.
The format is convenient for nuclear physicists who analyse experimental documentation and enter the summary of the experimental and data descriptions in EXFOR entries with the numerical data\footnote{
We remind the readers of difference in the production process of EXFOR and ENDF files.
The latter are not typed manually and not read by eyes, but written and read by computer programs (e.g., reaction model codes, processing codes).
}.
Compilation in the EXFOR format is supported by tools maintained by the data centres such as the EXFOR-Editor~\cite{Pikulina2024EXFOR} developed by Center of Nuclear Physics Data (Sarov) and the JANIS TRANS Checker~\cite{Soppera2014JANIS} developed by OECD NEA Data Bank (Paris).
HENDEL, another editor developed by Hokkaido University Nuclear Reaction Data Centre (Sapporo)~\cite{Otuka2002Development}, is designed for compilation by nuclear physicists who do not know the EXFOR format,
and graduate students and postdoctoral researchers of the university have drafted more than 500 EXFOR entries for data measured in Japan without noticing that they created EXFOR files.
The data compilation and exchange mechanisms are fully functional with the current EXFOR format,
and we do not see a strong reason to change the current EXFOR format for the compilation and exchange purpose.

The EXFOR format should not be confused with a centre-to-user format, namely a format tailored for end users.
A data file in the EXFOR format includes various abbreviations (e.g., CHSEP for chemical separation) and it makes reading of an EXFOR file bit difficult,
but it can be solved by development of an end user format.
For example,
the EXFOR web retrieval system maintained by the IAEA Nuclear Data Section~\cite{Zerkin2018Experimental} can display an EXFOR file in a more human readable style (``X4+'') for easy reading by eyes by inserting explanations to many EXFOR abbreviations,
and it is utilized for proofread of EXFOR entry drafts by the experimentalists who are not familiar with the EXFOR format.
The IAEA web retrieval system also provides the EXFOR information in the C4 format (Computation Format)~\cite{Cullen2001Program} where the EXFOR information is stored in a fixed set of units (i.e., all cross sections in b instead of in mb or $\mu$b) and fixed column order for better machine readability.
Another example of end user formats is the R33 format~\cite{Vickridge2003Update} developed for the excitation functions of the angular differential cross sections for the charged-particle induced reactions relevant to ion beam analysis.
Many EXFOR entries have been converted to the R33 format and disseminated through the IBANDL database~\cite{Dimitriou2017Nuclear}.
The data centres are encouraged to develop such end user formats to meet the needs of their customers.

Apart from these conversions for a selected part of the EXFOR library to other formats,
we start to see attempts ~\cite{Schnabel2020Computational,Lewis2023WPEC} to convert the full EXFOR library information into JSON (Java Script Object Notation), which key-value and hierarchical nature is suitable for accommodation of the information stored in the EXFOR format and also convenient for reading by some program languages like Python, which can read a JSON file very easily.
Schnabel~\cite{Schnabel2020Computational} developed and released an EXFOR-to-JSON converter and a tool to construct a MongoDB database from the JSON file.
Lewis et al.~\cite{Lewis2023WPEC} also discuss JSON as a choice to develop a database derived from the EXFOR library and supplemented by comments from users.
The IAEA EXFOR web retrieval system also provides JSON files in several JSON representations to end users.

Conversion of EXFOR to JSON could be useful not only for end users but also for data centres developing codes for compilation and dissemination of the EXFOR library.
Development of utility codes for editing, processing and checking EXFOR files may be simplified if they deal with the EXFOR information stored in JSON.
Various utility codes for EXFOR compilation and maintenance were developed in Fortran by US National Nuclear Data Center (NNDC) and are still used by EXFOR compilers.
Responsibility of their maintenance was transferred to the IAEA Nuclear Data Section (NDS) in 2000,
but maintenance of these Fortran codes becomes difficult after retirement of the Fortran experts.
Extension of the EXFOR format should not be constrained by presence of such legacy Fortran codes, and development of similar codes written in a programming language adopted by new generations is desired.
Furthermore, the NRDC network concluded in 2022 to support releasing all EXFOR codes as Open Source~\cite{Otuka2023Summary}.
Under the situation and movement,
we started development of new EXFOR utility codes widely adopting EXFOR files converted to JSON from the perspectives of both EXFOR compilers and users.

In this article,
we introduce newly developed utility codes \underline{for} \underline{EX}FOR Librar\underline{y} (ForEXy) utilizing a JSON representation of the EXFOR library.
It is followed by discussion on application of the codes to evaluation of the $^{237}$Np(n,f) cross section as an example.
Note that the latest versions of the ForEXy codes are freely available from the NRDC software website~\cite{NRDCSoftware} with a manual~\cite{Otuka2025ForEXy},
which also describes its installation from the PyPI Python package repository~\cite{PyPIForEXy} for use as modules.

\section{ForEXy utility codes}
\label{sec:ForEXy}

\begin{figure}[hbtp]
\centering
\includegraphics[bb=0 0 720 540,angle=0,width=1.0\linewidth]{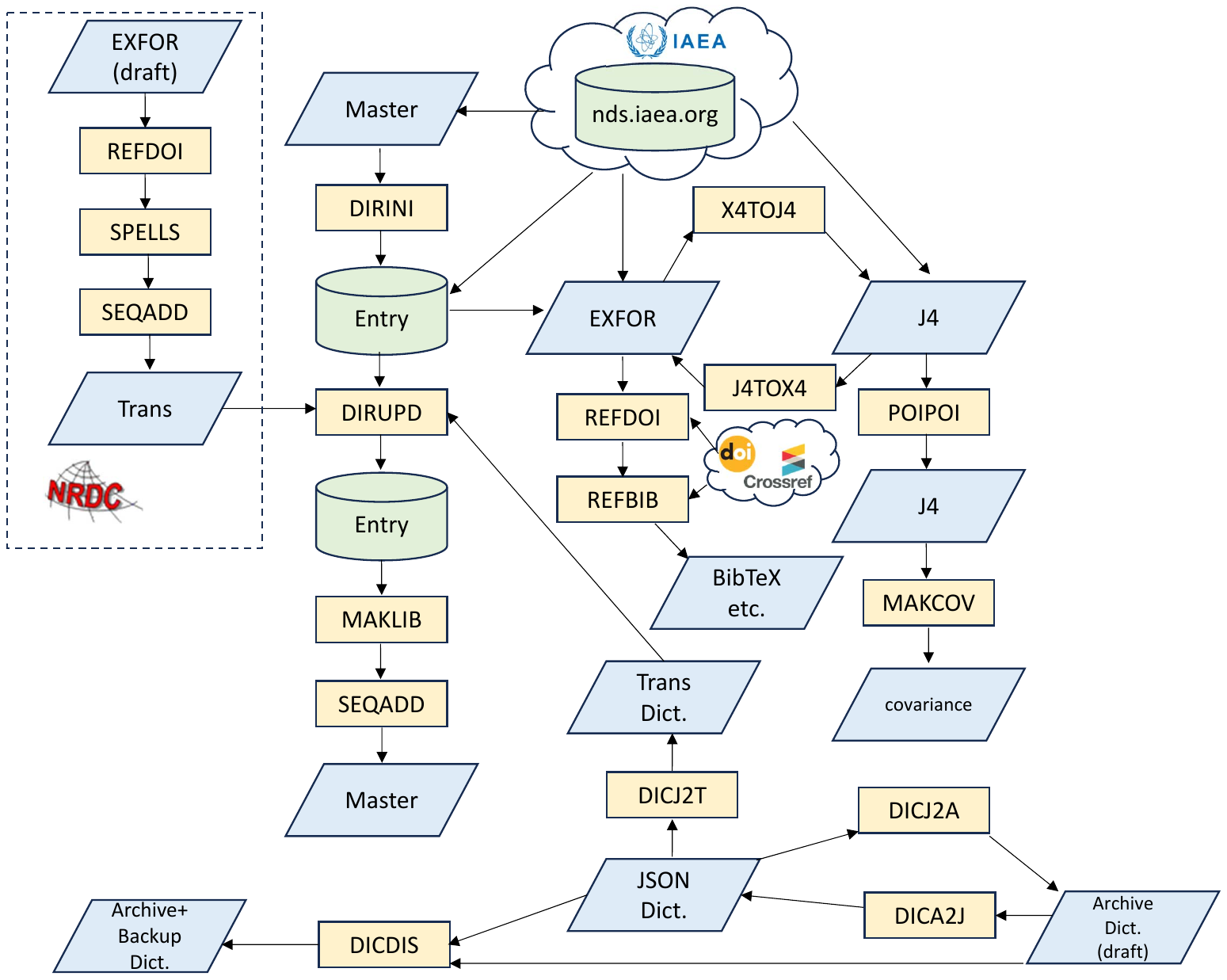}
\caption{
Flow of processing of EXFOR library and dictionary files by the ForEXy codes.
``Master'', ``Entry'' and ``Trans'' are the abbreviations of the EXFOR Master File, EXFOR Entry Files and EXFOR Trans File, respectively.
The JSON Dictionary is also used by J4TOX4, MAKCOV, POIPOI, REFBIB and X4TOJ4 but their use is not shown in the figure for better readability.
}
\label{fig:ForEXy}
\end{figure}

Table~\ref{tab:ForEXy} summarizes the seventeen utility codes included in the ForEXy package Ver.~2025.4.2.
All these codes are written in Python 3.
Three codes (REFBIB, REFDOI, SPELLS) require installation of external Python libraries, but the standard Python libraries are enough for other codes.
This section provides a brief introduction of each code.
Figure~\ref{fig:ForEXy} depicts the flow of EXFOR library and dictionary files processed by the ForEXy codes.

\begin{table}[hbtp]
\centering
\caption{
Codes included in ForEXy Ver.~2025.4.2.
}
\label{tab:ForEXy}
\begin{tabular}{ll}
\hline
\hline
\multicolumn{2}{l}{1. EXFOR storage maintenance}\\
\hline
DIRINI &Split an EXFOR library file (e.g., EXFOR Master File) into EXFOR entry files.\\
DIRUPD &Update EXFOR entry files with an EXFOR transmission (Trans) file.\\
MAKLIB &Merge EXFOR entry files into a single library file.\\
\hline
\multicolumn{2}{l}{2. Conversion between EXFOR and J4}\\
\hline
X4TOJ4 &Convert an EXFOR file to a J4 file.\\
J4TOX4 &Convert a J4 file to an EXFOR file.\\
\hline
\multicolumn{2}{l}{3. Pointer cancellation}\\
\hline
POIPOI &Cancel multiple reaction formalism in a J4 file.\\
EXTMUL &Extract a dataset compiled in the multiple reaction formalism in an EXFOR file.\\
\hline
\multicolumn{2}{l}{4. Covariance estimation}\\
\hline
MAKCOV &Produce a data table and covariance matrix from a J4 file.\\
\hline
\multicolumn{2}{l}{5. Bibliography management}\\
\hline
REFBIB &Produce bibliographies of references in an EXFOR file.\\
REFDOI &Obtain DOIs of references of an EXFOR file.\\
\hline
\multicolumn{2}{l}{6. Dictionary production}\\
\hline
DICA2J &Convert Archive Dictionaries to a JSON Dictionary.\\
DICJ2A &Convert a JSON Dictionary to Archive Dictionaries.\\
DICJ2T &Convert a JSON Dictionary to a Transmission (Trans) Dictionary.\\
DICDIS &Prepare Archive and Backup Dictionaries for distribution.\\
DIC227 &Produce Archive Dictionary 227 (nuclides) from a NUBASE file.\\
\hline
\multicolumn{2}{l}{7. EXFOR compilation support}\\
\hline
SPELLS &Check English spells in free text in an EXFOR file.\\
SEQADD &Add line sequence numbers to an EXFOR file.\\
\hline
\hline
\end{tabular}
\end{table}

\subsection{EXFOR storage maintenance (DIRINI, DIRUPD, MAKLIB)}
\label{sec:DIRINI}
The majority of the EXFOR users retrieve datasets by setting particular conditions such as the target nuclide (e.g., $^{56}$Fe), projectile (e.g., neutron) and quantity (e.g., cross section) on a web retrieval system, and download the data files of their interests to local storages.
However, some users may want to have global access to the EXFOR library (e.g., all (n,2n) reaction cross sections in EXFOR to benchmark an empirical formula or to train a machine learning tool).
To support such needs, the NRDC started discussion on public release of the EXFOR Master File, which is a single ASCII file containing all EXFOR entries. 

The EXFOR Master File was originally developed to resolve disagreement among the EXFOR files maintained by the data centres.
This disagreement was mainly caused by modifications to EXFOR files by other than the originating centre\footnote{
The originating centre means the data centre maintaining the EXFOR entry.
It is usually the data centre responsible for compilation of the data measured in the geographical area.
For example, US National Nuclear Data Center (NNDC) is usually the originating centre for an EXFOR entry compiling data measured in USA and Canada.
An EXFOR entry can be revised only by the originating centre of the EXFOR entry.
}.
The IAEA Nuclear Data Section analysed these differences, and created the first Common EXFOR Master File Ver.~2005-06-16, which was officially released on 1 July 2005~\cite{Schwerer2006Summary}.
This initial EXFOR Master File included unnecessary ``area 6'' entries (EXFOR entries made by conversion from the NNDC SCISRS files for old neutron induced reaction data measured in the NEA DB countries), and a revised Common EXFOR Master File (Ver.~2005-08-12) was released after elimination of the unnecessary entries.
Since then, the NRDC did not issue another EXFOR Master File for many years.

Considering potential needs of access to the full EXFOR library,
the NRDC concluded in 2023~\cite{Otuka2023Summary} to support update and release of the EXFOR Master File with assignment of a DOI (digital object identifiers) to each EXFOR Master File.
Accordingly, the NRDC started production and distribution of a new series of the EXFOR Master Files on the NRDC website~\cite{MasterFiles}.
The most recent one (EXFOR-2024) issued in December 2024~\cite{EXFOR2024} is a snapshot of the EXFOR library at the end of 2024.
The NRDC distributes this new series of the EXFOR Master Files under the terms of the CC BY 4.0 license~\cite{Otuka2024Summary}.
It means people can freely redistribute it in the original or another form, and also use it in a publication as long as the version (e.g., EXFOR-2024) is indicated and the EXFOR reference article~\cite{Otuka2014Towards} is cited.

About 500 new EXFOR entries are added to the EXFOR library every year, and we want to satisfy users who cannot wait the next release of the EXFOR Master File.
The NRDC requested the IAEA Nuclear Data Section to develop and release a tool generating the next EXFOR Maser File (e.g., EXFOR-2024) by updating the previous one (e.g., EXFOR-2023) with the EXFOR entries submitted in the current year (e.g., 2024)~\cite{Otuka2023Summary}.
The three tools DIRINI, DIRUPD and MAKLIB have been developed to satisfy these needs.

DIRINI is a tool to initialize the EXFOR files of a local storage.
Creation of a new storage under the directory \textit{entry} and its initialization with the EXFOR Master File Ver.~2023 can be done by
\begin{myleftbar}
%\begin{shadedf}
\begin{footnotesize}
\begin{verbatim}
x4_dirini.py -l exfor-2023.txt -d entry
\end{verbatim}
\end{footnotesize}
%\end{shadedf}
\end{myleftbar}
\noindent
where \textit{exfor-2023.txt} is downloaded from the NRDC EXFOR Master File website~\cite{MasterFiles} prior to the DIRINI operation.
When this operation completes,
we see an ASCII file of each EXFOR entry under a subdirectory structure (e.g., \textit{entry/1/10001.txt}, \textit{entry/1/10002.txt}, $\cdots$ \textit{entry/a/a0001.txt}, $\cdots$.).
Each subdirectory has a single character name corresponding to the first character of the entry number (area character), which characterizes the location of the measurement and projectile\footnote{
For example, ``1'' is for neutron induced reaction data from USA and Canada, and ``A'' for charged-particle induced reaction data from the former USSR countries. See~\cite{Otuka2022EXFOR} for a complete list of the areas.
}.

DIRUPD updates the local storage with the EXFOR entries released after production of the EXFOR Master File installed in the storage by DIRINI.
Each originating centre assembles newly created and revised EXFOR entries as a Trans file and submit it to other centres.
The Trans files are distributed from the EXFOR Trans File website~\cite{TransFiles}.
The storage initialized with EXFOR-2023 can be updated by
\begin{myleftbar}
\begin{footnotesize}
\begin{verbatim}
x4_dirupd.py -t trans.2318 -d entry
x4_dirupd.py -t trans.o098 -d entry
...
x4_dirupd.py -t trans.2319 -d entry
...
\end{verbatim}
\end{footnotesize}
\end{myleftbar}
\noindent
where \textit{trans.2318} and \textit{trans.o098} are the first and second EXFOR Trans files released in 2024.
This operation can be done for each area separately as long as it is done in chronological order for each area.
For example, one can load \textit{trans.2319} immediately after \textit{trans.2318}, but \textit{trans.o098} should be loaded before \textit{trans.o099}.

If we do not load Trans files in chronological order,
the latest version of an entry may be unintentionally overwritten by an older one.
For example,
the revision to EXFOR O0132 made in \textit{trans.o098} is cancelled if one loads \textit{trans.o096} (including an older version of this entry) after \textit{trans.o098} by mistake.
To prevent such a wrong operation,
DIRUPD compares the date and transmission ID written on the ENTRY record in the storage and the Trans file before updating the storage,
and print a warning message if the operation does not follow the chronological order.
For example,
DIRUPD prints the following warning messages if one tries to overwrite the EXFOR O0132 in the storage (copied from \textit{trans.o098}) by an older version in \textit{trans.o096}, and asks if the processing should continue:
\begin{myleftbar}
\begin{footnotesize}
\begin{verbatim}
** O0132.000: date in storage (20230615) > date in trans (20230519)
Continue? [Y] -->

** O0132.000: tape ID in storage (O098)  > tape ID in trans (O096)
Continue? [Y] -->
\end{verbatim}
\end{footnotesize}
\end{myleftbar}
Note that the most up-to-date set of the EXFOR entry files are distributed from the EXFOR Entry File website~\cite{EntryFiles} in a zipped form in the same directory structure,
and one can simply download and use it instead of performing the chain of the DIRUPD operations.

MAKLIB is a tool to merge the EXFOR entry files in the storage to a single ASCII file.
After completion of the DIRUPD operation for all Trans files released in 2024 in chronological order and do the same for the last dictionary released in 2024 (\textit{trans.9031}) saved as \textit{entry/9/90001.txt} in the storage,
one can reproduce the EXFOR Master File Ver.~2024 (\textit{exfor-2024.txt}) by the following operation:
\begin{myleftbar}
\begin{footnotesize}
\begin{verbatim}
x4_maklib.py -d entry -l library.txt -i 2024
x4_seqadd.py -i library.txt -o exfor-2024.txt
\end{verbatim}
\end{footnotesize}
\end{myleftbar}
\noindent
where the last operation (SEQADD) is just to print a line sequence number in every line.

\subsection{Conversion between EXFOR and J4 (X4TOJ4, J4TOX4)}
\label{sec:X4TOJ4}
As discussed in the introduction, several people have developed JSON representations of EXFOR according to their perspectives.
We designed our JSON representation of EXFOR (J4) by considering the following requirements:
\begin{itemize}
\item It is possible to reproduce the original EXFOR file from the JSON file.
\item It supports all logical structures of the EXFOR format.
\item It accommodates the value in each field of a code string with a particular key.
\item The meaning of each key in JSON is well defined.
\end{itemize}
Figure~\ref{fig:JSONcomp} compares two JSON representations for the decay data of $^{99m}$Tc adopted in an activation cross section measurement to discuss the second and third requirements.
The cross section determined with offline gamma spectroscopy is inversely proportional to the gamma intensity (gamma emission probability).
As the gamma intensity adopted in determination of the cross section may be revised later by a decay data expert, an activation cross section compiled in the EXFOR library may require renormalisation with the latest value of the gamma intensity.
Such manipulation can be simplified if the gamma intensity in EXFOR can be easily read by a computer program.
Both expressions in Fig.~\ref{fig:JSONcomp} are legal from the view of the JSON standard.
However, one cannot know from the first JSON representation which number gives the gamma intensity.
In contrast, the second JSON representation clearly indicates 140.5 and 0.885 are the energy and intensity of decay gammas, respectively.\footnote{
The radiation type DG stands for ``decay gammas'' as expanded in the EXFOR/CINDA Dictionary~\cite{Otuka2023EXFOR}.
We will discuss the Dictionary in Section~\ref{sec:DICA2J}.) 
}
The name of each key in the second example (flag, nuclide, half-life, radiation type, energy and intensity) is explained in the EXFOR Formats Manual~\cite{Otuka2022EXFOR},
which introduces that a code string under DECAY-DATA consists of three major fields ``nuclide'', ``half-life'', ``radiation'',
and the radiation field consists of three subfields ``radiation type'', ``energy'' and ``intensity''.

\begin{figure}[hbtp]
\begin{screen}
\begin{footnotesize}
\begin{verbatim}
"DECAY-DATA": {
  "coded_information": "(43-TC-99-M,6.0082HR,DG,140.5,0.885)"
},
\end{verbatim}
\end{footnotesize}
\end{screen}
\begin{screen}
\begin{footnotesize}
\begin{verbatim}
"DECAY-DATA": {
  "coded_information": {
    "flag": null,
    "nuclide": "43-TC-99-M",
    "half-life": {
      "value": 6.0082,
      "unit": "HR"
    },
    "radiation": {
      "radiation_type": "DG",
      "energy": 140.5,
      "intensity": 0.885
    }
  },
}
\end{verbatim}
\end{footnotesize}
\end{screen}
\caption{
Two JSON representations for the decay data of $^{99m}$Tc adopted in an activation cross section measurement ($T_{1/2}$=6.0082 h, $E_\gamma$=140.5~keV, $I_\gamma$=88.5\%)~\cite{Takacs2015Reexamination}. The second example is simplified from the actual J4 file of EXFOR D4322.002.
}
\label{fig:JSONcomp}
\end{figure}

We developed X4TOJ4 for conversion of an EXFOR file to a J4 file,
and J4TOX4 for conversion of a J4 file to an EXFOR file. 
X4TOJ4 reads an EXFOR file (e.g., \textit{exfor.txt}) containing a single EXFOR entry or several EXFOR entries,
and convert it to a J4 file (e.g., \textit{exfor.json}) by
\begin{myleftbar}
\begin{footnotesize}
\begin{verbatim}
x4_x4toj4.py -i exfor.txt -d dict_9131.json -o exfor.json
\end{verbatim}
\end{footnotesize}
\end{myleftbar}
\noindent
where \textit{dict\_9131.json} is an EXFOR/CINDA dictionary in JSON (will be discussed in Section~\ref{sec:DICA2J}).
A complete set of the J4 files produced from the latest EXFOR Entry Files are distributed from the NRDC website~\cite{J4Files}~\footnote{
This distribution is being done on a trial basis as of April 2025.
}.
J4TOX4 reads a J4 file and convert it to an EXFOR file, for example,
\begin{myleftbar}
\begin{footnotesize}
\begin{verbatim}
x4_j4tox4.py -i exfor.json -d dict_9131.json -o exfor.txt
x4_seqadd.py -i exfor.txt -o exfor_ord.txt
\end{verbatim}
\end{footnotesize}
\end{myleftbar}
\noindent
where the last operation (SEQADD) is just to print a line sequence number in every line.
We will discuss later why we need to process the J4TOX4 output by SEQADD.

Figure~\ref{fig:J4TOX4comp} shows an original EXFOR file and an EXFOR file converted from the original EXFOR file by X4TOJ4 and J4TOX4 for the EXFOR entry compiling a Mo($\gamma$,x)$^{93m}$Mo dataset~\cite{Timchenko2024Cross}.
Apart from the one column shift in the position of the name of the fifth author (V.A.~Kushnir),
we see differences in trailing zeros in real numbers, namely, ``739.50'' under DECAY-DATA and ``6.'' under ERR-3 in the original EXFOR file become ``739.5'' and ``6.0'' in the reproduced EXFOR file, respectively.
This is because X4TOJ4 treats a real number not as a string but as a real number in its default setting.
If we use ``-s'' option like
\begin{footnotesize}
\begin{verbatim}
x4toj4.py -i exfor.txt -d dict_9131.json -o exfor.json -s
\end{verbatim}
\end{footnotesize}
X4TOJ4 treats the real numbers in the original EXFOR file as characters, and the trailing zeros are preserved during the X4TOJ4 and J4TOX4 operations.

Another (minor) difference seen in Fig.~\ref{fig:J4TOX4comp} is the integers on the COMMON and ENDCOMMON records.
The second integer (N2) of the COMMON record and the first integer (N1) of the ENDCOMMON record give the number of the lines between the COMMON and ENDCOMMON records.
They are ``3'' in the original EXFOR file but are ``0'' in the EXFOR file produced by X4TOJ4 and J4TOX4.
Similar replacement is also seen in the second integer of the BIB record.
In general,
the integers (counters) in BIB (N2), ENDBIB (N1), COMMON (N2), ENDCOMMON (N1), ENDDATA (N1) and ENDSUBENT (N1) of an EXFOR file are set to zero in the corresponding J4 file (N1 and N2 mean the first and second integers of the record).
This is because these integers do not have meaning in the J4 file.
By this reason,
one needs to process a J4TOX4 output by SEQADD to replace the zeros of these integer fields with the correct integers.

\begin{figure}[hbtp]
\begin{screen}
\begin{footnotesize}
\begin{verbatim}
SUBENT        G4109001   20241110                                 
BIB                 14         35                                 
TITLE      Cross-sections of photoneutron reaction                
           natMo(gamma,xn)93mMo at the bremsstrahlung energy up   
           to 95 MeV                                              
AUTHOR     (I.S.Timchenko,O.S.Deiev,S.M.Olejnik,S.M.Potin,        
           V.A.Kushnir,V.V.Mytrochenko,S.A.Perezhogin)            
INSTITUTE  (4UKRKFT,3SLKSLO)                                      
REFERENCE  (J,VAT/I,,(3/151),15,2024)                             
           #doi:10.46813/2024-151-015                             
...
MONITOR    (42-MO-100(G,N)42-MO-99,,SIG,,BRA)                     
DECAY-MON  (42-MO-99,65.94HR,DG,739.50,0.1213)                    
ERR-ANALYS (ERR-T) A squared sum of statistical and systematic    
                   errors.                                        
...
COMMON               3          3                                 
ERR-1      ERR-2      ERR-3                                       
PER-CENT   PER-CENT   PER-CENT                                    
0.5        0.5        6.                                          
ENDCOMMON            3          0                                 
ENDSUBENT           42          0                                 
\end{verbatim}
\end{footnotesize}
\end{screen}
\begin{screen}
\begin{footnotesize}
\begin{verbatim}
SUBENT        G4109001   20241110                                 
BIB                 14          0                                 
TITLE      Cross-sections of photoneutron reaction                
           natMo(gamma,xn)93mMo at the bremsstrahlung energy up   
           to 95 MeV                                              
AUTHOR     (I.S.Timchenko, O.S.Deiev, S.M.Olejnik, S.M.Potin,     
            V.A.Kushnir, V.V.Mytrochenko, S.A.Perezhogin)         
INSTITUTE  (4UKRKFT,3SLKSLO)                                      
REFERENCE  (J,VAT/I,,(3/151),15,2024)                             
           #doi:10.46813/2024-151-015                             
...
MONITOR    (42-MO-100(G,N)42-MO-99,,SIG,,BRA)                     
DECAY-MON  (42-MO-99,65.94HR,DG,739.5,0.1213)                     
ERR-ANALYS (ERR-T) A squared sum of statistical and systematic    
                   errors.                                        
...
COMMON               3          0                                 
ERR-1      ERR-2      ERR-3                                       
PER-CENT   PER-CENT   PER-CENT                                    
0.5        0.5        6.0                                         
ENDCOMMON            0          0                                 
ENDSUBENT            0          0
\end{verbatim}
\end{footnotesize}
\end{screen}
\caption{
Original EXFOR file of G4109.001 (top) and the same subentry in an EXFOR file reproduced by processing the original one by X4TOJ4 without the ``-s'' option and J4TOX4 before processing by SEQADD (bottom).
The line sequence numbers at columns 67 to 79 are omitted for simplicity.
}
\label{fig:J4TOX4comp}
\end{figure}

J4TOX4 could open new possibility in EXFOR compilation.
One can create an EXFOR entry in J4 and its conversion to an EXFOR file by J4TOX4.
Unlike the EXFOR format,
 the J4 format is not suitable for manual editing.
But development of a J4 editor could overcome it,
and it could be easier than development of an EXFOR editor for those who are not familiar with the EXFOR format.

\subsection{Pointer cancellation (POIPOI, EXTMUL)}
\label{POIPOI}
The multiple reaction formalism of the EXFOR format allows to accommodate several datasets in a single subentry.
Figure~\ref{fig:multiple} shows that the $^{232}$Th(n,f)/$^{235}$U(n,f) and $^{232}$Th(n,f)/$^{238}$U(n,f) cross section ratios and absolute $^{232}$Th(n,f) cross section derived from the two cross section ratios are compiled in a single subentry.
The integer (``pointer'') 1, 2 or 3 at column 11 indicates that the information following the pointer describes the $^{232}$Th(n,f)/$^{235}$U(n,f) cross section ratio (pointer 1), $^{232}$Th(n,f)/$^{238}$U(n,f) cross section ratio (pointer 2) or absolute $^{232}$Th(n,f) cross section (pointer 3).
The use of the multiple reaction formalism can stress that these three datasets are mutually related to each other, and the J4 file produced from this EXFOR file by X4TOJ4 still preserves the same structure with the pointers since X4TOJ4 is designed to preserve the logical structure of the input EXFOR file.
However,
it has been known that pointers may introduce complications to development of a program processing EXFOR files. 

\begin{figure}[hbtp]
\begin{screen}
\begin{footnotesize}
\begin{verbatim}
----+----1----+----2----+----3----+----4----+----5----+----6----+-
SUBENT        23756002   20221007                                 
BIB                  7         19                                 
REACTION  1((90-TH-232(N,F),,SIG)/(92-U-235(N,F),,SIG))           
          2((90-TH-232(N,F),,SIG)/(92-U-238(N,F),,SIG))           
          3(90-TH-232(N,F),,SIG)                                  
...
ERR-ANALYS1(ERR-6,,,U) 232Th(n,f) counting statistics             
           (ERR-7,,,U) 235U(n,f) counting statistics              
          2(ERR-6,,,U) 232Th(n,f) counting statistics             
           (ERR-7,,,U) 238U(n,f) counting statistics              
          3(ERR-S)      Statistical uncertainty                   
           (MONIT1-ERR) 235U(n,f) cross section                   
           (MONIT2-ERR) 238U(n,f) cross section                   
...
ENDBIB              19          0                                 
NOCOMMON             0          0                                 
DATA                15          2                                 
EN         EN-RSL-HW  DATA      1ERR-6     1ERR-7     1DATA      2
ERR-6     2ERR-7     2DATA      3ERR-S     3ERR-4      MONIT1    3
MONIT1-ERR3MONIT2    3MONIT2-ERR3                                 
MEV        MEV        NO-DIM     PER-CENT   PER-CENT   NO-DIM     
PER-CENT   PER-CENT   B          PER-CENT   PER-CENT   B          
PER-CENT   PER-CENT   PER-CENT                                    
 2.0        0.1        0.0908     3.4        1.1        0.218     
 3.2        1.2        0.117      4.3        0.         1.289     
 1.3        0.538      1.3                                        
 2.5        0.1        0.0887     3.0        1.0        0.205     
 2.9        1.1        0.112      2.7        0.         1.261     
 1.3        0.545      1.3                                        
ENDDATA             12          0                                 
ENDSUBENT           36          0                                 
\end{verbatim}
\end{footnotesize}
\end{screen}
\caption{
EXFOR file of 23756.002 (original) compiling three datasets (1) $^{232}$Th(n,f)/$^{235}$U(n,f) cross section ratio, (2) $^{232}$Th(n,f)/$^{238}$U(n,f) cross section ratio, and (3) absolute $^{232}$Th(n,f) cross section measured in an experiment~\cite{Michalopoulou2021Measurement}.
The line sequence numbers at columns 67 to 79 are omitted for simplicity.
}
\label{fig:multiple}
\end{figure}

POIPOI is a code to extract a single dataset from several datasets compiled together in the same subentry. Extraction of the absolute $^{232}$Th(n,f) cross section in an EXFOR file including EXFOR 23756.002 (\textit{exfor.txt}\footnote{
This file must include EXFOR 23756.001 in addition to 002.
}
) to a J4 file (\textit{exfor\_poi.json}) can be done by 
\begin{myleftbar}
\begin{footnotesize}
\begin{verbatim}
x4_x4toj4.py -i exfor.txt -d dict_9131.json -o exfor.json
x4_poipoi.py -i exfor.json -d dict_9131.json -e 23756.002.3\
             -o exfor_poi.json
\end{verbatim}
\end{footnotesize}
\end{myleftbar}
\noindent
where the backslash on the second line indicates that the arguments of the second command (POIPOI) continues to the third line.
The EXFOR file is first converted to a J4 file \textit{exfor.json} by X4TOJ4, and then the information belonging to the $^{232}$Th(n,f) cross section (pointer 3) in the J4 file is extracted by POIPOI.
\begin{figure}[hbtp]
\begin{screen}
\begin{footnotesize}
\begin{verbatim}
...
"SUBENT": {
  "N1": "23756002",
  "N2": 20221007,
  ...
},
...
"DATA": {
  ...
  "pointer": [
    "",          "",          "1",         "1",         "1",         "2",
    "2",         "2",         "3",         "3",         "",          "3",
    "3",         "3",         "3"
  ],
  "heading": [
    "EN",        "EN-RSL-HW", "DATA",      "ERR-6",     "ERR-7",     "DATA", 
    "ERR-6",     "ERR-7",     "DATA",      "ERR-S",     "ERR-4",     "MONIT1",
    "MONIT1-ERR","MONIT2",    "MONIT2-ERR"
  ],
  "unit": [
    "MEV",       "MEV",       "NO-DIM",    "PER-CENT",  "PER-CENT",  "NO-DIM",
    "PER-CENT",  "PER-CENT",  "B",         "PER-CENT",  "PER-CENT",  "B",
    "PER-CENT",  "PER-CENT",  "PER-CENT"
  ],
  "value": [
    [
    2.0,         0.1,         0.0908,      3.4,         1.1,         0.218,
    3.2,         1.2,         0.117,       4.3,         0.0,         1.289,
    1.3,         0.538,       1.3
    ],
    [
    2.5,         0.1,         0.0887,      3.0,         1.0,         0.205,
    2.9,         1.1,         0.112,       2.7,         0.0,         1.261,
    1.3,         0.545,       1.3
    ]
  ]
},
...
\end{verbatim}
\end{footnotesize}
\end{screen}
\caption{
J4 file of 23756.002 converted from an EXFOR file including EXFOR 23756.002 (Fig.~\ref{fig:multiple}) by X4TOJ4.
}
\label{fig:beforepoipoi}
\end{figure}
\begin{figure}[hbtp]
\begin{screen}
\begin{footnotesize}
\begin{verbatim}
...
"SUBENT": {
  "N1": "23756002",
  "N2": 20221007,
  ...
},
...
"DATA": {
  ...
  "pointer": [
    "3",         "3",         "3",         "3",         "3",         "3",         
    "3",         "3",         "3"
  ],
  "heading": [
    "EN",        "EN-RSL-HW", "DATA",      "ERR-S",     "ERR-4",     "MONIT1",
    "MONIT1-ERR","MONIT2",    "MONIT2-ERR"
  ],
  "unit": [
    "MEV",       "MEV",       "B",         "PER-CENT",  "PER-CENT",  "B",
    "PER-CENT",  "PER-CENT",  "PER-CENT"
  ],
  "value": [
    [
    2.0,         0.1,         0.117,       4.3,         0.0,         1.289,
    1.3,         0.538,       1.3
    ],
    [
    2.5,         0.1,         0.112,       2.7,         0.0,         1.261,
    1.3,         0.545,       1.3
    ]
  ]
},
...
\end{verbatim}
\end{footnotesize}
\end{screen}
\caption{
J4 file of 23756.002.3 extracted from the J4 file of 23756.002 (Fig.~\ref{fig:beforepoipoi}) by POIPOI.
}
\label{fig:afterpoipoi}
\end{figure}
Figure~\ref{fig:beforepoipoi} shows a J4 file converted from the EXFOR file of 23756.002 (Fig.~\ref{fig:multiple}) by X4TOJ4,
and Fig.~\ref{fig:afterpoipoi} shows a J4 file extracted from the J4 file of 23756.002 (Fig.~\ref{fig:beforepoipoi}) for the pointer 3 by POIPOI.
We observe in Fig.~\ref{fig:afterpoipoi} that the J4 file processed by POIPOI collects the numbers relevant to the pointer 3 (i.e., absolute $^{232}$Th(n,f) cross section) of the EXFOR file.

EXTMUL is a code to extract a dataset in the multiple reaction formalism in an EXFOR file by running X4TOJ4, POIPOI and J4TOX4 sequentially.
It (1) converts the original EXFOR file to a J4 file by X4TOJ4,
(2) extracts a dataset in the J4 file to a secondary J4 file by POIPOI,
and (3) finally converts the secondary J4 file to the EXFOR format by J4TOX4.
Extraction of the absolute $^{232}$Th(n,f) cross section in the original EXFOR file (\textit{exfor.txt}) to a derived EXFOR file (\textit{exfor\_poi.txt}) can be done by
\begin{myleftbar}
\begin{footnotesize}
\begin{verbatim}
x4_extmul.py -i exfor.txt -d dict_9131.json -e 23756.002.3\
             -o exfor_poi.txt
x4_seqadd.py -i exfor_poi.txt -o exfor_ord.txt
\end{verbatim}
\end{footnotesize}
\end{myleftbar}
\noindent
where the second operation (SEQADD) is just to print a line sequence number in every line. The result of the original EXFOR file (Fig.~\ref{fig:multiple}) processed by EXTMUL is shown in Fig.~\ref{fig:nonmultiple}.
The three characters ``???'' in the SUBENT record of the EXTMUL output signifies that this subentry does not include the full information of the original EXFOR 23756.002 dataset anymore.

\begin{figure}[hbtp]
\begin{screen}
\begin{footnotesize}
\begin{verbatim}
SUBENT        23756???   20221007                                 
BIB                  7          0                                 
REACTION   (90-TH-232(N,F),,SIG)                                  
...
ERR-ANALYS (ERR-S)      Statistical uncertainty                   
           (MONIT1-ERR) 235U(n,f) cross section                   
           (MONIT2-ERR) 238U(n,f) cross section                   
...
ENDBIB               0          0                                 
NOCOMMON             0          0                                 
DATA                 9          2                                 
EN         EN-RSL-HW  DATA       ERR-S      ERR-4      MONIT1     
MONIT1-ERR MONIT2     MONIT2-ERR                                  
MEV        MEV        B          PER-CENT   PER-CENT   B          
PER-CENT   PER-CENT   PER-CENT                                    
 2.0        0.1        0.117      4.3        0.         1.289     
 1.3        0.538      1.3                                        
 2.5        0.1        0.112      2.7        0.         1.261     
 1.3        0.545      1.3                                        
ENDDATA              0          0                                 
ENDSUBENT            0          0                                 
\end{verbatim}
\end{footnotesize}
\end{screen}
\caption{
Absolute $^{232}$Th(n,f) cross section in EXFOR 23756.002.3 extracted from the original EXFOR file (Fig.~\ref{fig:multiple}) by EXTMUL before processing by SEQADD.
}
\label{fig:nonmultiple}
\end{figure}

\subsection{Covariance estimation (MAKCOV)}
\label{sec:MAKCOV}
Some nuclear applications require not only the best estimate of the quantity of interest but also its uncertainty and even covariance.
Major evaluated nuclear data libraries tend to include more and more covariances to satisfy such nuclear data needs.
Its importance has been known for low energy neutron induced reactions in the relation with fission energy application for many decades,
but we also start to see needs of uncertainties in charged-particle induced reaction data for evaluation of the beam monitor cross sections~\cite{Hermanne2018Reference}.
It is obvious that a certain fraction of the covariances in the evaluated nuclear data originates from the covariances in experimental nuclear data, and hence it is important to archive the experimental reaction data in the EXFOR library with the associated covariance information.
As many experimentalists are not familiar with the concept of covariance, some guides have been published to encourage the experimentalists to provide the covariance related information along with the best estimate~\cite{Mannhart2013Small,Smith2012Experimental,Becker2012Data,Otuka2017Uncertainty,Neudecker2023Templates}.
The EXFOR format has been extended to accommodate such information received from the experimentalists~\cite{Otuka2012Experimental},
and a web tool was developed to construct a covariance matrix from the information~\cite{Zerkin2012Web}.

Suppose a quantity $y$ is expressed by products and/or quotients of $n$ independent parameters.
The total fractional covariance (\%$^2$) between two quantities $y_i$ and $y_j$ is expressed in terms of $n$ partial fractional covariances by
\begin{equation}
\cov(y_i,y_j) = \cov^1(y_i,y_j) + \cov^2(y_i,y_j)+ \cdots + \cov^n(y_i,y_j),
\label{eqn:covprop}
\end{equation}
where $\cov^k(y_i,y_j)$ ($k=1,n$) is the partial fractional covariance (\%$^2$) between $y_i$ and $y_j$ due to the uncertainty in the $k$th parameter.
Appendix B of the report on the $^{241}$Am(n,2n)$^{240}$Am activation cross section measurement performed at JRC IRMM~\cite{Sage2017Mesures} nicely demonstrates numerical examples of this equation with a detailed explanation on each term.
However, such covariance information is hardly documented by experimentalists, and the evaluators usually should estimate it based on the information available in the publication and EXFOR entry.
A practical way to estimate the covariance is to modify Eq.~(\ref{eqn:covprop}) to
\begin{equation}
\cov(y_i,y_j) = C^1_{ij}\, \Delta^1y_i\, \Delta^1 y_j + C^2_{ij}\, \Delta^2 y_i\, \Delta^2y_j+\cdots + C^n_{ij}\, \Delta^ny_i\, \Delta^n y_j,
\label{eqn:covpropmod}
\end{equation}
where $C^k_{ij}$ ($k=1,n$) is the correlation coefficient between $i$ and $j$ due to the uncertainty in the $k$th parameter ($0 \le C^k_{ij} \le 1$, $C^k_{ii}=1$) and $\Delta^k y_i$ is the partial fractional uncertainty (\%) in $y_i$ due to the uncertainty in the $k$th parameter.
For example,
$C^k_{ij}=\delta_{ij}$ (Kronecker's delta) if $k$ is for a number of counts, 
and $C^k_{ij}=1$ if $k$ is for an overall normalisation factor.
By setting $i=j$, Eq.~(\ref{eqn:covpropmod}) becomes the well-known quadrature sum formula to calculate the square of the total fractional uncertainty as the sum of the squares of the partial fractional uncertainties:
\begin{equation}
\var (y_i) = \left(\Delta^1y_i\right)^2 + \left(\Delta^2y_i\right)^2+ \cdots +\left(\Delta^ny_i\right)^2.
\label{eqn:uncprop}
\end{equation}
In an EXFOR file, the total uncertainty is coded under the heading ERR-T and partial uncertainties are coded under the headings ERR-S (statistical uncertainty), ERR-1 (first partial uncertainty), ERR-2 (second partial uncertainty) and so on.

The tool MAKCOV reads a J4 file processed by POIPOI,
and calculates the correlation coefficients $\cor(y_i,y_j)=\cov (y_i,y_j)/\sqrt{\var(y_i) \var(y_j)}$ and the total uncertainty $\sqrt{\var(y_i)}$ following Eqs.~(\ref{eqn:covpropmod}) and (\ref{eqn:uncprop}), respectively.
The correlation coefficient $C^k_{ij}$ in Eq.~(\ref{eqn:covpropmod}) is usually not provided by the experimentalists,
and evaluators must estimate it based on the documentation and EXFOR entry.
MAKCOV reads an extra input file (``HED'' file) defining the headings of the independent and dependent variables as well as the correlation coefficients for the partial uncertainties.
See an appendix of the ForEXy manual~\cite{Otuka2025ForEXy} for details on the syntax of the HED file.

Figure~\ref{fig:makcovinp} shows an EXFOR file of the absolute $^{235}$U(n,f) cross section~\cite{Wasson1982Absolute}.
If one wants to construct the correlation coefficients from the three partial uncertainties coded under the headings ERR-S, ERR-1 and ERR-2, the HED file may consist of the following nine lines:
\begin{footnotesize}
\begin{verbatim}
1: EN-MIN
2: EN-MAX
3:
4: DATA
5: 
6: 
7: ERR-S      0.
8: ERR-1      1.
9: ERR-2      1.
\end{verbatim}
\end{footnotesize}
where the last three lines instruct MAKCOV that the correlation coefficient is 0 for the uncertainty coded under ERR-S while it is 1 for those coded under ERR-1 and ERR-2.

\begin{figure}[hbtp]
\begin{screen}
\begin{footnotesize}
\begin{verbatim}
SUBENT        10595002   20170510                                 
BIB                  8         20                                 
REACTION   (92-U-235(N,F),,SIG,,AV)                               
...
ERR-ANALYS (ERR-T) Total uncertainty                              
           (ERR-S) Statistical uncertainty                        
           (ERR-1) Shape systematic uncertainty                   
           (ERR-2) Normalization uncertainty                      
...
ENDBIB              20          0                                 
COMMON               4          3                                 
EN-NRM-MIN EN-NRM-MAX MONIT      ERR-2                            
KEV        KEV        B          PER-CENT                         
 10.       20.         2.55       2.4                             
ENDCOMMON            3          0                                 
DATA                 6         77                                 
EN-MIN     EN-MAX     DATA       ERR-T      ERR-S      ERR-1      
KEV        KEV        B          PER-CENT   PER-CENT   PER-CENT   
   5.0        5.5      3.987      3.7        1.5        2.4       
   5.5        6.0      4.171      3.7        1.6        2.3       
   6.0        6.5      3.445      3.7        1.7        2.2       
   6.5        7.0      3.284      3.6        1.8        2.1       
...
\end{verbatim}
\end{footnotesize}
\end{screen}
\caption{
$^{235}$U(n,f) absolute cross section compiled in EXFOR 10595.002 with the total and partial uncertainties.
The line sequence numbers at columns 67 to 79 are omitted for simplicity.
}
\label{fig:makcovinp}
\end{figure}

By using this HED file (\textit{exfor\_hed.txt}) and the EXFOR file (\textit{exfor.txt}\footnote{
This file must include EXFOR 10595.001 in addition to 002.
}),
one can obtain the correlation coefficients of this dataset by the following operation:
\begin{myleftbar}
\begin{footnotesize}
\begin{verbatim}
x4_x4toj4.py -i exfor.txt -d dict_9131.json -o exfor.json
x4_poipoi.py -i exfor.json -d dict_9131.json -e 10595.002\
             -o exfor_poi.json
x4_makcov.py -i exfor_poi.json -j exfor_hed.txt -d dict_9131.json\
             -e 10595.002 -o x4_makcov_out.txt
\end{verbatim}
\end{footnotesize}
\end{myleftbar}
\noindent
It shows the EXFOR file is first converted to a J4 file \textit{exfor.json} by X4TOJ4,
and further processed by POIPOI to obtain a secondary J4 file \textit{exfor\_poi.json} before processing by MAKCOV.

Figure~\ref{fig:makcovout} shows the final output (\textit{x4\_makcov\_out.txt}).
A table with four columns (incident energy, half of incident energy bin width, cross section and its total uncertainty) for 77 data points in EXFOR 10595.002 is followed by their correlation coefficients.

In many occasions,
users just need a cross section table without correlation coefficients.
By replacing the sixth to ninth lines of the HED file with
\begin{footnotesize}
\begin{verbatim}
6:            +
7: ERR-S      0.
8: ERR-1      0.
9: ERR-2      0.
\end{verbatim}
\end{footnotesize}
one can obtain the four-column table including the total uncertainty calculated from the three partial uncertainties according to Eq.~(\ref{eqn:uncprop}) without the correlation coefficient. Namely, MAKCOV can be also used as a calculator of the total uncertainty\footnote{
This operation is of course unnecessary for the EXFOR 10595.002 case since the EXFOR file provides the total uncertainty under ERR-T.
One can check if the ERR-T values can be reproduced from the ERR-S, ERR-1 and ERR-2 values by MAKCOV.
}.

Application of MAKCOV to our cross section evaluation work will be presented in Section~\ref{sec:evaluation}.

\begin{figure}[hbtp]
\begin{screen}
\begin{footnotesize}
\begin{verbatim}
#10595.002   O.A.Wasson,1976
#20170510    1USANBS                      77
#x          dx         y          dy
#eV         eV         b          b
 5.2500E+03 2.5000E+02 3.9870E+00 1.4795E-01
 5.7500E+03 2.5000E+02 4.1710E+00 1.5388E-01
 6.2500E+03 2.5000E+02 3.4450E+00 1.2653E-01
 6.7500E+03 2.5000E+02 3.2840E+00 1.2026E-01
...

# cor(x,y) estimated with absolute dataset treatment
 1.000
 0.824 1.000
 0.810 0.799 1.000
 0.795 0.784 0.772 1.000
...
\end{verbatim}
\end{footnotesize}
\end{screen}
\caption{
Cross section table and correlation coefficients of the $^{235}$U(n,f) cross section in EXFOR 10595.002 printed by MAKCOV.
}
\label{fig:makcovout}
\end{figure}

\subsection{Bibliography management (REFBIB, REFDOI)}
\label{sec:REFDOI}
EXFOR users should cite the references (source articles) of the EXFOR entries used in their publications.
However, preparation of proper citations is time-consuming work in general.
EXFOR compilers are instructed to provide the title and author list of the primary reference (the reference coded in the first line of REFERENCE) under the keyword TITLE and AUTHOR.
However, the title and author list under these keywords may be those edited by the EXFOR compiler when the EXFOR entry contains datasets from several final publications, and none of the final publications provide the title and author list covering the whole contents of the EXFOR entry.
The EXFOR system also does not define a rule to express characters not in the ASCII character set (e.g., accented characters, Greek characters).
It means a citation list prepared by extraction from the TITLE and AUTHOR lines of EXFOR files must be often corrected by comparison with the source articles.
One could suggest to the NRDC improvement of the contents of these bibliography keywords so that they can be directly usable for citations.
However, compilation of the bibliographies is not a primary goal of the EXFOR library.
There is a bibliography database dedicated to nuclear physics publications (NSR~\cite{Pritychenko2011Nuclear}),
and the EXFOR compilers should not spend much time to enrich the bibliographies in the EXFOR library.
We developed REFBIB to support EXFOR users for preparation of a citation list from EXFOR files under this situation.

REFBIB provides the bibliographies deposited in CrossRef, which is the most important agency for registration of DOIs for journal articles.
Their publishers register in the CrossRef database not only the DOIs but also the metadata (including bibliographies) of their journal articles,
and the deposited metadata are very useful for citation preparation.
REFBIB returns the bibliographies in various formats (e.g., BibTeX) if it finds that the journal volume is registered in the CrossRef database.

The following REFBIB command line produces a BibTeX entry for the article coded as J,JPJ,6,66,1951~\cite{Arakatsu1951Photo} in EXFOR K2180:
\begin{myleftbar}
\begin{footnotesize}
\begin{verbatim}
x4_refbib.py -i J,JPJ,6,66,1951 -d dict_9131.json -o refbib.bib\
             -r bibtex -e email@address.com
\end{verbatim}
\end{footnotesize}
\end{myleftbar}
\noindent
which prints in \textit{refbib.bib} the BibTeX entry shown in the upper part of Fig.~\ref{fig:bibtexout}.
Note that REFBIB also processes a DOI of a journal article,
and it can be useful for researchers other than the EXFOR users.
The lower part of Fig.~\ref{fig:bibtexout} shows an output printed by the following REFBIB command line for an article not dealing with nuclear science~\cite{Kroesbergen1976Harpsichord}.
\begin{myleftbar}
\begin{footnotesize}
\begin{verbatim}
x4_refbib.py -i 10.1093/earlyj/4.4.439 -d dict_9131.json\
             -o refbib.bib -r bibtex -e email@address.com
\end{verbatim}
\end{footnotesize}
\end{myleftbar}
\noindent
\begin{figure}[hbtp]
\begin{screen}
\begin{footnotesize}
\begin{verbatim}
@article{,
  author  ={Arakatsu, Bunsaku and Sonoda, Masateru and Uemura, Yoshiaki and ...
  title   ={The photo-disintegration of be by the high energy ...
  journal ={Journal of the Physical Society of Japan},
  year    ={1951},
  volume  ={6},
  pages   ={66--67},
  doi     ={10.1143/JPSJ.6.66}
}
\end{verbatim}
\end{footnotesize}
\end{screen}
\begin{screen}
\begin{footnotesize}
\begin{verbatim}
@article{,
  author  ={Kroesbergen, Willem and Koopman, Ton},
  title   ={Harpsichord building in {Holland}},
  journal ={Early Music},
  year    ={1976},
  volume  ={4},
  pages   ={439--442},
  doi     ={10.1093/earlyj/4.4.439}
}
\end{verbatim}
\end{footnotesize}
\end{screen}
\caption{
BibTeX entries printed by REFBIB for a nuclear physics article~\cite{Arakatsu1951Photo} (upper panel) and for an article irrelevant to nuclear science~\cite{Kroesbergen1976Harpsichord} (bottom panel).
}
\label{fig:bibtexout}
\end{figure}

REFDOI outputs the DOIs of all journal article reference codes in an EXFOR file.
This code converts the EXFOR file to a J4 file, checks presence of each journal article in the CrossRef database by REFBIB, and finally prints a list of DOIs.
This tool can be used by EXFOR compilers as a reference code validator.
If the CrossRef database does not return a DOI,
it could imply presence of a typo in the EXFOR reference code or an error in the metadata of the CrossRef database.

To examine the performance of REFDOI,
we extracted all journal articles coded as the primary references from the EXFOR Master File Ver.~2024 (EXFOR-2024) and processed them by REFDOI.
We were able to identify DOIs in the CrossRef database for 82\% of all (22554) primary references from journals.
This exercise concluded that 14\% of them are not registered in the CrossRef database, and they are mainly those published in Russian journals such as Yadernaya Fizika (YF), Izvestiya Rossiiskoi Akademii Nauk (IZV) and Atomnaya Energiya (AE).

Figure~\ref{fig:REFDOIout} shows an output of REFDOI (\textit{x4\_refdoi\_out.txt}) obtained for an EXFOR file (\textit{exfor.txt}) by
\begin{myleftbar}
\begin{footnotesize}
\begin{verbatim}
x4_refdoi.py -i exfor.txt -d dict_9131.json -o x4_refdoi_out.txt\
             -e email@address.com 
\end{verbatim}
\end{footnotesize}
\end{myleftbar}
\noindent
It shows that REFDOI could not receive a DOI for the second reference code (J,JP/CS,940,012005,2017)~\cite{Gyuerky2018Nuclear}.

\begin{figure}[hbtp]
\begin{screen}
\begin{footnotesize}
\begin{verbatim}
D4319.001 REFERENCE J,PR/C,90,052801,2014   10.1103/PhysRevC.90.052801
D4319.001 REFERENCE J,JP/CS,940,012005,2017 ** Suspicious reference code
\end{verbatim}
\end{footnotesize}
\end{screen}
\caption{
Output of REFDOI for an EXFOR file containing EXFOR D4319.001. It is known that the publisher initially printed 2017 as the publication year of the second reference~\cite{Gyuerky2018Nuclear}, and later corrected it to 2018.
}
\label{fig:REFDOIout}
\end{figure}

\subsection{Dictionary production (DICA2J, DICJ2A, DICJ2T, DICDIS, DIC227)}
\label{sec:DICA2J}
The EXFOR/CINDA Dictionary~\cite{Otuka2023EXFOR} is a set of dictionaries, which are the ``heart'' of the EXFOR system.
The Dictionary has several roles.
First,
it provides the expansions (explanations) of codes (abbreviations) used in EXFOR files.
For example,
the code ``ARI'' is expanded to ``Applied Radiation and Isotopes'' in Dictionary 5 (Journals).
Second,
it gives various flags ensuring the EXFOR compilation rule and supporting compilation and dissemination tools.
For example, the code ``,SIG'' (cross section) is defined in Dictionary 236 (Quantities) with a flag ``B'', which indicates that this quantity must be always combined with a unit code defined in the Dictionary 25 (Data units) with the same flag ``B'' (e.g., B, MB and MICRO-B).
The flag ``B'' (cross section) itself is also defined in Dictionary 26 (Unit family codes).

The IAEA Nuclear Data Section updates the Dictionary twice a year for the NRDC,
and distributes it in three different forms officially – Archive, Backup and Trans.
For every update of the Dictionary, the Archive Dictionaries are edited manually, while the Backup and Trans Dictionaries are converted from the Archive Dictionaries.
Note that the Backup and Trans Dictionaries do not carry the full Dictionary information from the Archive Dictionaries, and therefore the conversion procedure is not reversible.

This conversion was done by several Fortran codes written by NNDC till 2023.
As maintenance of these Fortran codes becomes difficult for the current incumbents of the IAEA Nuclear Data Section,
we developed new Python codes to produce Backup and Trans Dictionaries from the Archive Dictionaries.
DICA2J converts the Archive Dictionaries to JSON.
This JSON Dictionary is a single ASCII file containing the full dictionary information, and we can reproduce the Archive Dictionaries from the JSON Dictionary by using DICJ2A.
The Trans Dictionary is also a single ASCII file, and it can be produced from the JSON Dictionary by DICJ2T.
Finally, the Backup Dictionary (also a single ASCII file) is produced from the Archive Dictionaries by DICDIS.
DICDIS also prepares Archive Dictionaries for distribution by removing the flags given for editorial purposes only.

Among the Archive Dictionaries, Dictionary 227 (Nuclides) is produced from the NUBASE file~\cite{Kondev2021Nubase2020},
which is currently updated every four to five years by Argonne National Laboratory (USA) and Institute of Modern Physics (China).
When a new NUBASE file is released,
we convert it to the Archive Dictionary format by using DIC227 and use it for production of the Backup and Trans Dictionaries.

We developed the JSON Dictionary to make programming for conversion from the Archive Dictionaries to the Backup and Trans Dictionaries easier.
But the JSON Dictionary could also be useful for EXFOR software developers and end users.
We designed the JSON Dictionary by considering the requirements discussed at the beginning of Section~\ref{sec:X4TOJ4}.
Figure~\ref{fig:DSONcomp} compares two possible JSON representations for definition of the quantity code ``PAR,SIG'' (partial cross section~\footnote{
In EXFOR,
the partial cross section means a cross section characterized by a secondary energy.
The cross section for excitation of the residual nucleus to a particular level and the cross section for production of gammas at particular energy are examples of the partial cross section.
}).
In the first JSON representation,
one cannot know the meaning of the two additional codes ``CSP'' and ``B''.
Contrary,
the keys seen in the second JSON representation (``code'', ``reaction type code'', ``unit family code'', ``expansion'') are well described in the EXFOR/CINDA Dictionary Manual~\cite{Otuka2023EXFOR},
and it helps proper use of all values in the dictionary entry.
As seen in the preceding subsections,
this JSON Dictionary (e.g., \textit{dict\_9131.json}) is utilized in many other tools of the ForEXy package.

\begin{figure}[hbtp]
\begin{screen}
\begin{footnotesize}
\begin{verbatim}
{
  "code": "PAR,SIG",
  "additional_code_1": "CSP",
  "additional_code_2": "B",
  "description": "Partial cross section",
},
\end{verbatim}
\end{footnotesize}
\end{screen}
\begin{screen}
\begin{footnotesize}
\begin{verbatim}
{
  "code": "PAR,SIG",
  "reaction_type_code": "CSP",
  "unit_family_code": "B",
  "expansion": "Partial cross section",
},
\end{verbatim}
\end{footnotesize}
\end{screen}
\caption{
Two JSON representations for the quantity code ``PAR,SIG'' (partial cross section). The second example is simplified from an entry of the real JSON Dictionary.
}
\label{fig:DSONcomp}
\end{figure}

The following series of operations
(1) generates the JSON Dictionary Ver.~9131 in the directory \textit{json} from the Archive Dictionaries in the directory \textit{input} by DICA2J,
(2) generates the Trans Dictionary Ver.~9131 in the directory \textit{dist} from the JSON Dictionary in the directory \textit{json} by DICJ2T,
and
(3) generates the Archive Dictionaries and Backup Dictionary Ver.~9131 for distribution in the directory \textit{dist} from the Archive Dictionaries in the directory \textit{input} and the JSON Dictionary in the directory \textit{json} by DICDIS:
\begin{myleftbar}
\begin{footnotesize}
\begin{verbatim}
x4_dica2j.py -n 9131 -i input -o json
x4_dicj2t.py -n 9131 -i json -o dist
x4_dicdis.py -n 9131 -a input -j json -o dist
\end{verbatim}
\end{footnotesize}
\end{myleftbar}
\noindent
Note that the input and output file names are hardcoded in these tools.

\subsection{Compilation support (SPELLS, SEQADD)}
\label{sec:SPELLS}
The free text description part of the EXFOR library is not designed for machine processing, and its importance for EXFOR end users is not as high as the part designed for machine processing.
From the view of quality assurance of the library, however, it is desirable to exclude typos in the free text.
The free text in the EXFOR library is in English.
However, it is not efficient to apply a generic English spell checker to EXFOR files.
This is because the files contain many EXFOR codes (abbreviations) which generic English spell checkers do not understand.
Also various technical terms (e.g., linac, epithermal) and abbreviations (e.g., mb, mg/cm2) should not be detected as typos in EXFOR spell checking.

SPELLS is a tool to check English spells in an EXFOR file.
It extracts free text from an EXFOR file, and detects English typos by using pyspellchecker, an external Python spell checker available in the PyPI Python package repository.
It also reads an external dictionary collecting technical terms and abbreviations to accept them during spell checking.
Checking of an EXFOR file \textit{exfor.txt} with an external dictionary \textit{x4\_spells.dic}\footnote{
See the ForEXy manual~\cite{Otuka2025ForEXy} for an example of the external dictionary.
}can be performed by
\begin{myleftbar}
\begin{footnotesize}
\begin{verbatim}
x4_spells.py -i exfor.txt -d x4_spells.dic -o x4_spells_out.txt
\end{verbatim}
\end{footnotesize}
\end{myleftbar}
\noindent
, which prints a report in \textit{x4\_spells\_out.txt}.

SEQADD edits an EXFOR file by adding a line sequence number on the right side (columns 67 to 80) of each line,
and also by updating the integers following some system identifiers such as BIB and ENDBIB records.

\section{Application of ForEXy codes to cross section evaluation}
\label{sec:evaluation}
Simulation with a particle transport code requires cross section at any incident energy and, at the same time, the cross section must be unique at the energy.
Cross section evaluation interpolates experimental data points to obtain the cross section as a continuous function of the incident energy.
When there are few experimental data points, this interpolation requires a code implementing reaction models such as the Hauser-Feshbach statistical model~\cite{Hauser1952Inelastic}.
On the other hand, fitting without a physics model may be an adequate evaluation approach when enough experimental data points are available.
In the JENDL project (e.g.,~\cite{Iwamoto2023Japanese}),
neutron induced fission cross sections of many target nuclides have been evaluated with least-squares fitting codes such as SOK~\cite{Kawano2000Simultaneous} and GMA~\cite{Poenitz1997Simultaneous}.
We have recently performed evaluation for the neutron induced fission cross sections of $^{232}$Th, $^{233,235,238}$U and $^{239,240,241,242}$Pu~\cite{Devi2024EXFOR,Otuka2022EXFORbased,Okuyama2024EXFOR} based on the experimental fission cross sections and their ratios in the EXFOR library with the SOK code.
We have also recently extended this approach to evaluation of the $^{233}$U neutron capture cross section~\cite{Otuka2025Evaluation} by using the experimental fission-to-capture cross section ratio (alpha value) of $^{233}$U by the SOK code.
In this section,
we first briefly summarize the formalism adopted in the SOK code,
and then introduce our ongoing evaluation work for the $^{237}$Np neutron fission cross section with the SOK and ForEXy codes.

\subsection{Simultaneous evaluation with SOK}
\label{sec:sok}
Suppose that we perform fitting of $n$ evaluated cross sections to $m$ experimental data points assuming that
\begin{equation}
\mathbf{y} = C \mathbf{x} + \mathbf{e},
\end{equation}
where $\mathbf{y}$ is a $m$ dimensional experimental data point vector,
$\mathbf{x}$ is a $n$ dimensional evaluated cross section vector,
and $\mathbf{e}$ is a $m$ dimensional vector corresponding to the fitting residual.
The $n\times m$ matrix $C=\{c_{ij}\}$ is known as the design matrix (sensitivity coefficients).
In our evaluation,
we model the relationship between $\mathbf{y}$ and $\mathbf{x}$ by
\begin{equation}
c_{ij}=\left\{
\begin{array}{ll}
(\epsilon_i - E_{j-1}) / (E_j - E_{j-1}) &\text{if } E_{j-1} \le \epsilon_i < E_j     \\
(\epsilon_i - E_{j+1}) / (E_j - E_{j+1}) &\text{if } E_j     \le \epsilon_i < E_{j+1} \\
0                                        &\text{otherwise,}
\end{array}
\right.
\end{equation}
where $\epsilon_i$ gives the incident energy of the experimental data point $y_i$,
while $E_j$ gives the incident energy (energy grid) of the evaluated cross section $x_j$.
If $\epsilon_i$ is the middle between $E_j$ and $E_{j+1}$,
this model gives $y_i = (x_j + x_{j+1}) / 2 + e_i$.
Our problem is to determine $\mathbf{x}$ for a given $\mathbf{y}$.
The least-squares solution $\mathbf{x}$ and its $n\times n$ covariance $X$ are
\begin{eqnarray}
\mathbf{x} &=& X C^t V^{-1} \mathbf{y} 
\label{eqn:GLSQsol1}
\\
X          &=& (C^tV^{-1}C)^{-1},
\label{eqn:GLSQsol2}
\end{eqnarray}
where $V$ is the $m\times m$ covariance matrix of the experimental data point vector $\mathbf{y}$.
In the SOK code,
the vectors $\mathbf{x}$ and $\mathbf{y}$ are the logarithms of cross sections for several target nuclides and/or reaction channels (e.g., $^{235}$U(n,f) and $^{239}$Pu(n,f)) and $\mathbf{y}$ may also include the logarithms of their ratios (e.g., $^{239}$Pu(n,f)/ $^{235}$U(n,f)).
This logarithmic transformation linearizes the cross section ratios for inclusion in the least-squares approach,
and allows us to evaluate cross sections for various reactions simultaneously without converting the cross section ratios to absolute cross sections.
See Ref.~\cite{Kawano2000Simultaneous} for more details on the formulation and~\cite{Otuka2023Simultaneous} for a schematic explanation of the formalism.

The SOK code obtains the least-squares solution not by calculating Eqs.~(\ref{eqn:GLSQsol1}) and (\ref{eqn:GLSQsol2}) but by updating a prior estimate (evaluated cross sections before update) $\mathbf{x_0}$ and its covariance $X_0$ to a posterior estimate (evaluated cross sections after update) $\mathbf{x_1}$ and its covariance $X_1$ with the following equations~\cite{Kawano2000Simultaneous,Hirtz2024Parameter}:
\begin{eqnarray}
\mathbf{x_1} &=& \mathbf{x_0} + X_0 C^t (CX_0C^t + V)^{-1} (\mathbf{y} - C\mathbf{x_0})\\
X_1          &=& X_0 - X_0 C^t (CX_0C^t + V)^{-1} C X_0,
\end{eqnarray}
where $\mathbf{y}$ and $V$ are the experimental data point vector and its covariance matrix used in this update step. 

The most important part of the evaluation with least-squares fitting is proper estimation of $V$.
As discussed in Section 2.4, very few experimental works provide this matrix, and usually the evaluator should estimate it by considering the data reduction procedure and uncertainty quantification mentioned in the article and EXFOR entry.
Figure 2 of~\cite{Otuka2022EXFORbased} demonstrates that the fitting result may be very sensitive to $V$, and it is worth to spend time for its estimation.

It is also time consuming to extract partial uncertainties from EXFOR entries and to construct the covariances according to Eq.~(\ref{eqn:covpropmod}) if we have to do it manually.
We automatised this process by developing a code SOX (\underline{SO}K input from E\underline{X}FOR),
which reads EXFOR entries and prints $\mathbf{y}$ and $V$.
This tool was originally written in Perl and used in our previous simultaneous fission cross section evaluations.
As construction of $\mathbf{y}$ and $V$ not from EXFOR files but from J4 files makes the source code of the SOX code more traceable,
we completely rewrote the SOX code by adopting Python and J4.
The new SOX code processes EXFOR files by X4TOJ4, POIPOI and MAKCOV to obtain $\mathbf{y}$ and $V$,
and prints them following the SOK input format specification.

\subsection{Evaluation of $^{237}$Np(n,f) cross section}
\label{sec:np237}
As discussed above,
the SOK code allows us to include an experimental cross section ratio in $\mathbf{y}$,
and we prefer to use the ratio without conversion to the absolute cross section.
The reference reaction appearing in the denominator of the ratio is typically a fission like $^{235,238}$U(n,f) or light nuclide reaction like $^1$H(n,n)$^1$H, $^6$Li(n,t)$^4$He and $^{10}$B(n,$\alpha$)$^7$Li reactions,
for which the IAEA Neutron Data Standards~\cite{Carlson2018Evaluation} provide the standard cross section values determined by a simultaneous evaluation with the GMA code.
It is not trivial to include all reference reactions in our simultaneous evaluation framework by several reasons,
and 
we include the ratios converted to the absolute cross sections in our framework for the cross sections measured relative to the light nuclide standard reaction cross sections.

In contrast to these light nuclide standard reactions,
the number of experimental datasets is not too large for $^{237}$Np(n,f),
which is considered as an alternative to $^{238}$U(n,f) as a standard in the MeV region~\cite{Cierjacks1977237Np},
and some cross sections in our experimental database dedicated to the simultaneous evaluation are measured relative to the $^{237}$Np(n,f) cross section.
Therefore,
we are currently working on adding $^{237}$Np(n,f) in our framework so that we can include the relevant cross section ratios (e.g., $^{235}$U(n,f)/$^{237}$Np(n,f)) in our evaluation framework without conversion to the absolute cross sections (e.g., $^{235}$U(n,f)).

We set the lower and upper boundary to 100~keV and 200~MeV, respectively, for the $^{237}$Np(n,f) cross section evaluation,
and extended the experimental database originally constructed for $^{233,235,238}$U and $^{239,240,241}$Pu simultaneous fission cross section evaluation~\cite{Otuka2022EXFORbased2} by adding $^{237}$Np(n,f) related experimental datasets in this energy range.
Table~\ref{tab:exp237Np} summarizes the number of experimental datasets included in our evaluation by April 2025.

\begin{table}[hbtp]
\centering
\caption{
Numbers of the experimental $^{237}$Np fission datasets added in simultaneous evaluation. ``$^{237}$Np/$^{235}$U'' means the $^{237}$Np(n,f)/$^{235}$U(n,f) fission cross section ratio.
}
\label{tab:exp237Np}
\begin{tabular}{lcccccc}
\hline
\hline
Reaction             &
$^{237}$Np           &
$^{237}$Np/$^{235}$U   &
$^{237}$Np/$^{238}$U   &
$^{237}$Np/$^{239}$Pu  &
$^{235}$U/$^{237}$Np  &
$^{238}$U/$^{237}$Np \\
\hline
Datasets    &
10 &
15 &
1  &
2  &
1  &
3  \\
\hline
\hline
\end{tabular}
\end{table}

To confirm that the original SOX (Perl) and rewritten SOX (Python) implement the construction of $\mathbf{y}$ and $V$ in the same way,
we obtained $\mathbf{y}$ and $V$ (or more precisely, the total uncertainty and correlation coefficient) from the EXFOR files by the Perl and Python versions.
As expected,
the Python version requires more CPU time for the processing since it converts EXFOR files to J4 files before starting calculation of the total uncertainty and correlation coefficients.
We found that $\mathbf{y}$ from the two versions perfectly agree,
but found slight differences in $V$, mainly for those constructed for the $^{240}$Pu(n,f)/$^{235}$U(n,f) cross section ratio dataset measured by Behrens et al. (EXFOR 10597.002~\cite{Behrens1978Measurements}).

\begin{figure}[hbtp]
\centering
\includegraphics[angle=0,width=1.0\linewidth, bb=0 0 842 595]{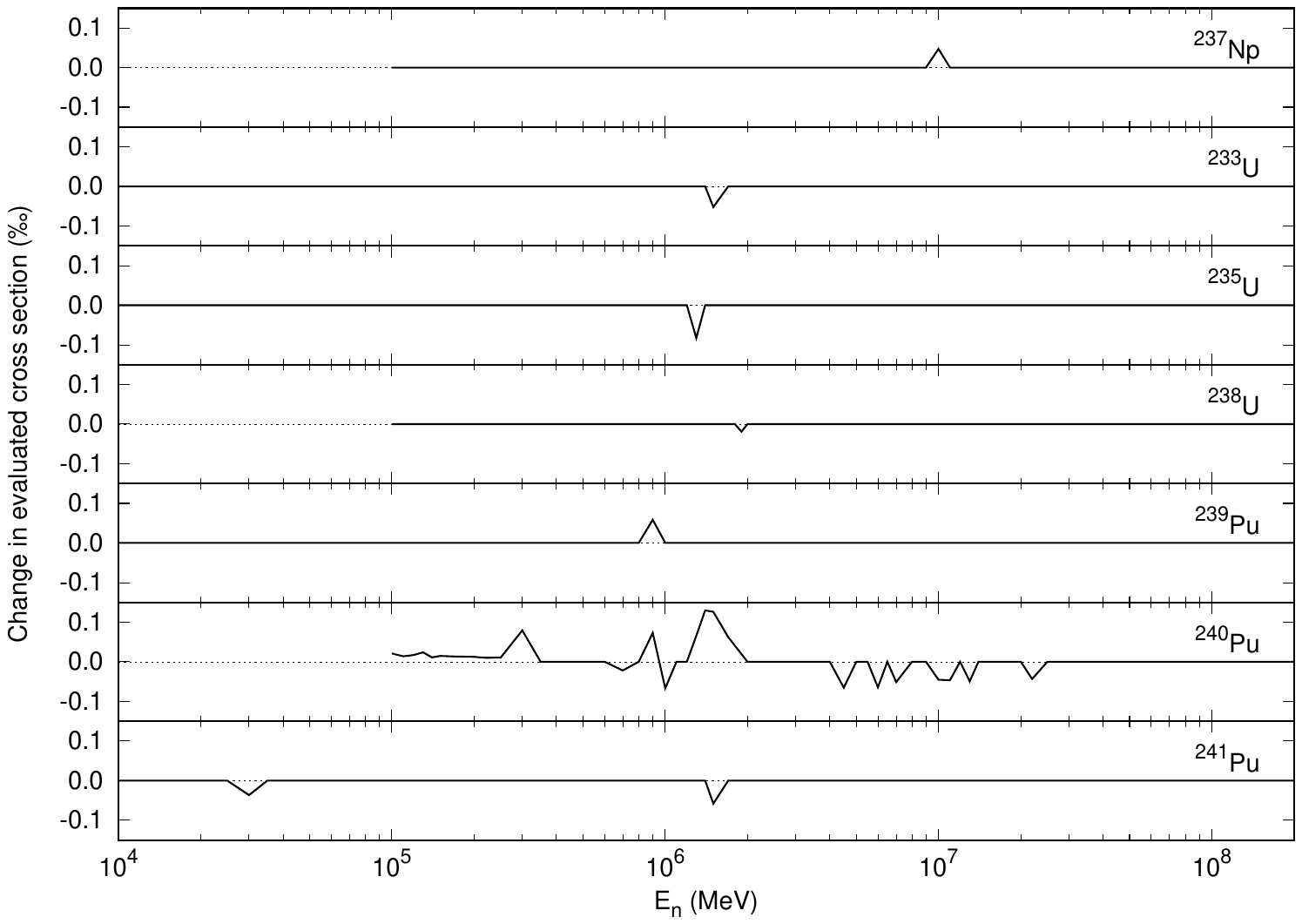}
\caption{
Change in $^{237}$Np, $^{233,235,238}$U, $^{239,240,241}$Pu(n,f) cross sections evaluated by SOK due to replacement of SOX (Perl) with SOX (Python).
}
\label{fig:SOXcomp}
\end{figure}

To see the impact of the replacement of the SOX version,
we performed fitting by the SOK code with $\mathbf{y}$ and $V$ obtained by both Perl and Python versions.
Figure~\ref{fig:SOXcomp} shows the relative difference $(\sigma_\text{Python} - \sigma_\text{Perl})/\sigma_\text{Perl}$ in the evaluated $^{237}$Np, $^{233,235,238}$U, $^{239,240,241}$Pu(n,f) cross sections $\mathbf{x}$.
It shows the largest change in the $^{240}$Pu(n,f) cross section but it is less than 0.1\textperthousand.
We therefore decided to move from SOX (Perl) to SOX (Python) for production of $\mathbf{y}$ and $V$\} from EXFOR files.

\begin{figure}[hbtp]
\centering
\includegraphics[angle=0,width=1.0\linewidth, bb=0 0 842 595]{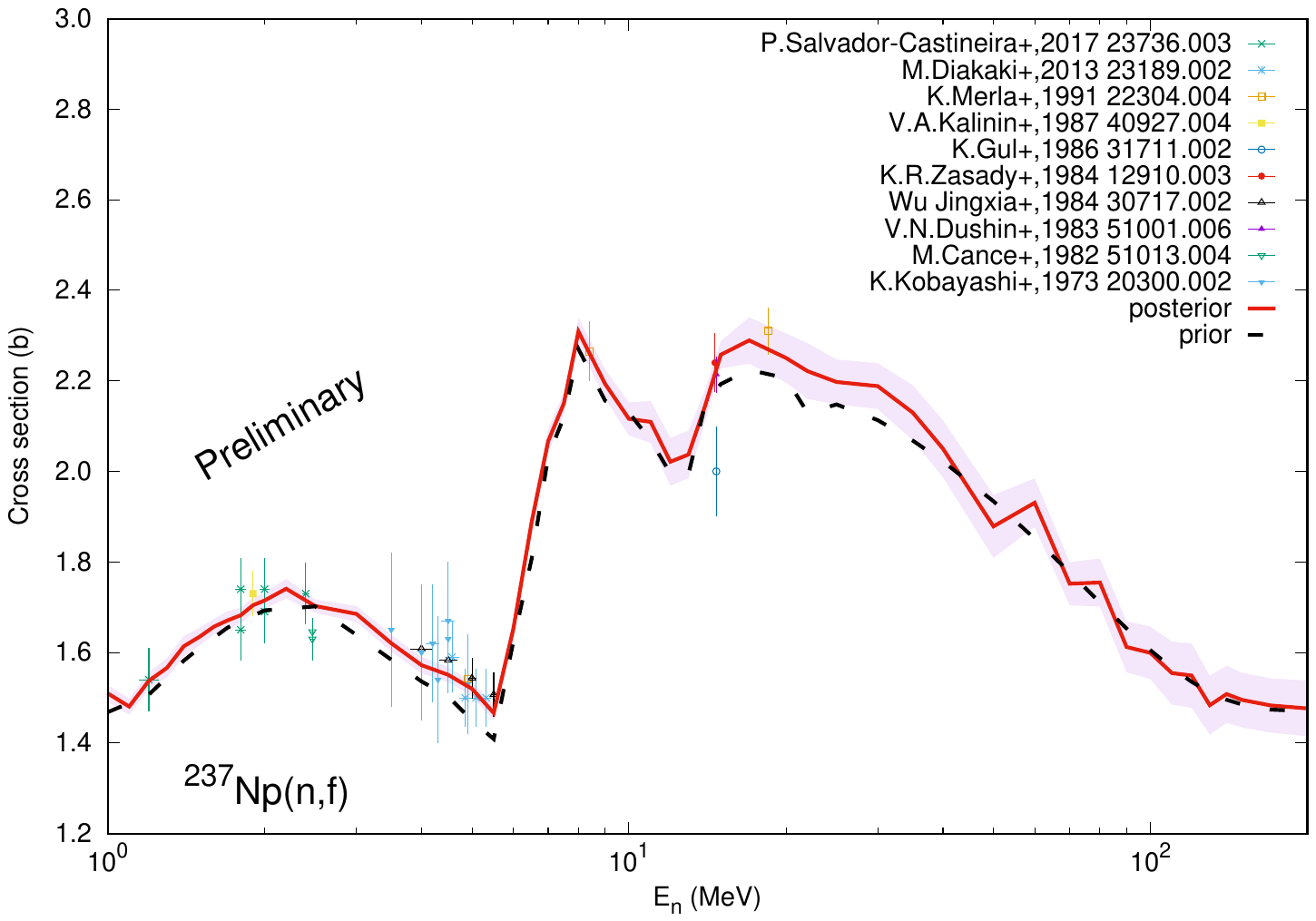}
\caption{
Comparison of the prior and posterior values with the experimental data points~\cite{
Salvador-Castineira2017Absolute,
Diakaki2013Determination,
Merla1991Absolute,
Kalinin1987Absolute,
Gul1986Measurements,
Zasadny1984Measurement,
Wu1984Measurement,
Dushin1983Statistical,
Cance1982Mesures,
Kobayashi1973Measurement}
used in the simultaneous fitting for the $^{237}$Np(n,f) cross section (preliminary result).
}
\label{fig:np237}
\end{figure}

\begin{figure}[hbtp]
\centering
\includegraphics[angle=0,width=1.0\linewidth, bb=0 0 842 595]{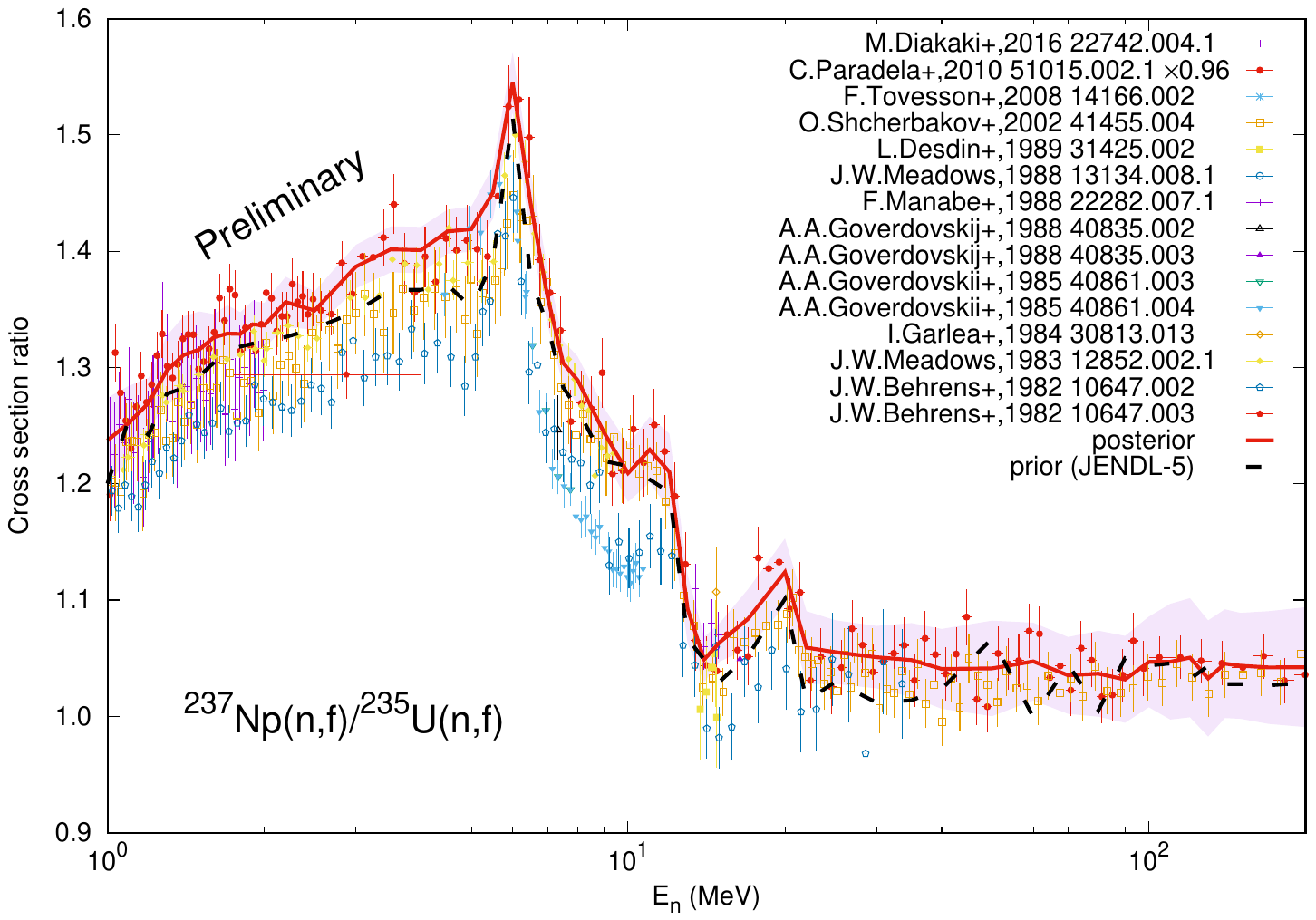}
\caption{
Comparison of the prior and posterior values with the experimental data points~\cite{
Diakaki2016Neutron,
Paradela2010Neutron,
Tovesson2008Subthreshold,
Shcherbakov2002Neutron,
Desdin1989Measurement,
Meadows1988Fission,
Goverdovskij1988Measurement,
Goverdovskii1985Measurement,
Garlea1984Measuring,
Meadows1983Fission,
Behrens1982Measurement}
used in the simultaneous fitting for the $^{237}$Np(n,f)/ $^{235}$U(n,f) cross section ratio (preliminary result).
}
\label{fig:np237u235}
\end{figure}

Figures~\ref{fig:np237} and \ref{fig:np237u235} compare the prior and posterior evaluated values with the experimental data points adopted in the fitting procedure for the $^{237}$Np(n,f) absolute cross section and $^{237}$Np(n,f)/ $^{235}$U(n,f) cross section ratio~\footnote{
An EXFOR dataset ID starting from 5 (e.g., 51001.006) in the figure legends indicates that the original EXFOR entry modified by us is used in the present evaluation.
See Sect.3.3 of~\cite{Otuka2022EXFORbased} for further details.
}.
The prior values of the $^{237}$Np(n,f) and $^{235}$U(n,f) cross sections are taken from the JENDL-5 library.
Note that the JENDL-5 $^{237}$Np(n,f) cross section in the fast neutron region below 20~MeV is taken from the JENDL-4.0 library~\cite{Shibata2011JENDL}, which does not consider the experimental datasets published after Shcherbakov et al.~\cite{Shcherbakov2002Neutron}.

During trial fitting,
we realized that Paradela et al.'s $^{237}$Np(n,f)/ $^{235}$U(n,f) cross section ratio~\cite{Paradela2010Neutron} is systematically higher than the ratios from other measurements.
Therefore,
we treated Paradela et al.'s ratio as a ``shape ratio'' in the final fitting.
It means we treated their absolute ratio in the EXFOR library as a ratio in arbitrary units, and determined the overall normalisation factor of the ratio as an additional fitting parameter.
Paradela et al.'s ratio shown in Fig.~\ref{fig:np237u235} is the original one multiplied by the overall normalisation factor determined in our fitting (0.96).
After multiplication of this normalisation factor, Paradela et al.'s ratio is still systematically higher than Shcherbakov et al.'s ratio, which is followed well by the JENDL-5 evaluation.

If we compare the $^{252}$Cf spontaneous fission neutron spectrum averaged cross section (SACS), the JENDL-5 SACS (1347 mb) is  lower than the SACS recommended by Mannhart (1361 mb)~\cite{Mannhart2006Response}, and this supports the systematic increase from the prior to the posterior seen in Fig.~\ref{fig:np237}.
Very recently, Vorobyev et al. published a new $^{237}$Np(n,f)/$^{235}$U(n,f) ratio dataset between 0.1 and 400~MeV~\cite{Vorobyev2024Measurement}, and we plan to check if the fitting including this new dataset still gives the $^{237}$Np(n,f) cross section systematically higher than the JENDL-5 cross section once the newly published ratio is compiled by the Russia Nuclear Data Centre (Obninsk) and becomes available in the EXFOR library.

\section{Summary}
\label{sec:summary}
We designed a JSON representation of the EXFOR library and developed a converter from an EXFOR file to a JSON file.
The latest EXFOR entries in the JSON format (J4) are distributed from the NRDC website on a trial basis,
but one can create and update the same J4 files by using the ForEXy codes and EXFOR files freely available from the NRDC website.
We also developed a set of the tools utilizing the J4 files to support EXFOR compilers, software engineers and end users.
Adoption of JSON drastically simplified reading of the EXFOR information, and it should help people who want to develop their own codes dealing with the EXFOR library.

It has been known that the EXFOR format is convenient for EXFOR compilation but not convenient for users who want to read EXFOR files by writing computer programs.
The X4TOJ4 converter clearly separates these two contradicting needs and satisfy compilers and users with keeping the EXFOR format for the compilation and exchange purpose.
The data files and codes introduced in this article are released under CC BY 4.0 and MIT license, respectively,
and we anticipate that their release stimulates development of useful databases and tools by those who are not very familiar with the EXFOR format.

We believe that development and improvement of the IT infrastructure for EXFOR compilation and dissemination are beneficial for the community.
For example,
EXFOR users can obtain a cross section table (e.g., Fig.~\ref{fig:makcovout}) from an EXFOR file very easily with MAKCOV.
However, it does not solve the main bottleneck in the flow of the information from experimentalists to users.
Katharine Way et al. define compilation in scientific fields as follows~\cite{Way1969Foreward}:
\begin{quote}
To compile, says Webster,
is ``to compose out of material from other documents.''
But in scientific fields the juxtaposition of values obtained by different authors for a given quantity is of little use unless the values have first been made comparable by making sure they are based on the same standards, include the same corrections,
present uncertainties calculated in the same way, etc.
Thus the compiler's first task is to produce this \textit{comparability} or to note the impossibility of doing so in special cases.
By a ``compilation'' of scientific data we will mean a collection of results of different authors in which the values have been made comparable.
\end{quote}
To make the EXFOR library as a collection of comparable datasets,
EXFOR compilers have to (1) select an experimental dataset useful for comparison from an article,
(2) tag the dataset with a reaction/quantity identifier defined in the EXFOR dictionary and rule (REACTION coding) in spite of the variety in the nomenclature and symbols,
and (3) collect from the experimentalist the numerical data including the reference data (e.g., decay data, standard or monitor value) adopted in the data reduction process.\footnote{
To ensure comparability,
one has to renormalise the originally published dataset with the latest reference data.
The EXFOR library stores originally published datasets instead of renormalised ones,
but keeps the reference data adopted by the experimentalist for future renormalisation.
In this regard,
tools renormalising datasets in the EXFOR library based on the original and latest reference data (e.g., X4Pro~\cite{Zerkin2023X4Pro}) would be useful.
}
It is obvious that a nuclear physicist can start reasonable EXFOR compilation only after training given by an experienced EXFOR compiler and also only after establishing good relationship with the experimentalists of the region.
The community should allocate resource to sustain access to the experimental data and their comparability over the next decades.
The Asian NRDC members (CNDC, JAEA, JCPRG, KNDC, NDPCI) periodically organise regional workshops in cooperation with the IAEA to enhance capabilities of EXFOR compilation in their region~\cite{Takibayev2014Fourth,Saxena2015Fifth,Sarsembayeva2015Sixth,Chen2016Seventh,Odsuren2017Eigth,Yang2018Ninth,Zholdybayev2019Tenth},
and NDPCI also organises the biennial workshop to transfer the knowledge of EXFOR to young nuclear physicists in India.
As discussed in Sec.~\ref{sec:intro},
there is an editor allowing researchers to make EXFOR files without knowing the EXFOR format,
and formatting is not the most challenging part in creation of an EXFOR entry.

We applied the ForEXy codes to evaluation of the $^{237}$Np(n,f) cross section between 100 keV and 200 MeV within the simultaneous evaluation framework including $^{233,235,238}$U and $^{239,240,241}$Pu(n,f) cross sections.
The preliminary result shows the newly evaluated $^{237}$Np(n,f) cross section is systematically higher than the one compiled in the JENDL-5 library.
We continue our evaluation by adding newly measured datasets including the $^{237}$Np(n,f)/$^{235}$U(n,f) cross section ratio dataset recently published by Vorobyev et al.~\cite{Vorobyev2024Measurement}.

\section*{Acknowledgement}
We received valuable comments on the new development from David Brown (US National Nuclear Data Center), Oscar Cabellos (Universidad Politécnica de Madrid), Yongli Jin (China Nuclear Data Center), Boris Pritychenko (US National Nuclear Data Center), Georg Schnabel (IAEA Nuclear Data Section) and Nicolas Soppera (OECD NEA Data Bank).
Ludmila Marian (IAEA Nuclear Data Section) helped formulation of the statement allowing CC BY 4.0 license and DOI assignment for the NRDC products such as the EXFOR Master File.
We are deeply grateful to the reviewers for their careful reading of this manuscript and also testing some of the ForEXy tools.
We also would like to thank the NRDC members for their continuous dedication to the EXFOR library maintenance and the experimentalists those who enrich the contents of the EXFOR library by submission of their numerical experimental data.
%\appendix
%\section{Example Appendix Section}
%\label{app1}
%Appendix text.

\bibliography{np237-ANP2024.bib}
\end{document}